\documentclass[conference, final]{IEEEtran}
\IEEEoverridecommandlockouts
\usepackage{color}
\usepackage{soul}
\usepackage{cite}
\usepackage[acronym,shortcuts]{glossaries}
\usepackage{amsmath,amssymb,amsfonts}
\usepackage{algorithmic}
\usepackage{graphicx}
\usepackage{textcomp}
\usepackage{xcolor}
\usepackage{siunitx}
\def\BibTeX{{\rm B\kern-.05em{\sc i\kern-.025em b}\kern-.08em
    T\kern-.1667em\lower.7ex\hbox{E}\kern-.125emX}}
\usepackage{array,multirow}
\usepackage{hyperref}
\newcommand{\parag}[1]{\noindent\textbf{#1. }}
\usepackage{todonotes}

\usepackage{cleveref}   
\newacronym{adsb}{ADS-B}{Automatic Dependant Surveillance-Broadcast}
\newacronym{gnss}{GNSS}{Global Navigation Satellite System}
\newacronym{ml}{ML}{Machine Learning}
\newacronym{mlat}{MLAT}{Multilateration}
\newacronym{sdr}{SDR}{Software Defined Radio}
\newacronym{tdoa}{TDoA}{Time Difference Of Arrival}
\newacronym{cots}{COTS}{commercial off-the-shelf}
\newacronym{svm}{SVM}{Support Vector Machine}

\usepackage{array}
\newcolumntype{L}[1]{>{\raggedright}m{#1}}
\newcolumntype{R}[1]{>{\raggedleft}m{#1}}
    
\begin{document}

\title{LoVe is in the Air -- Location Verification of ADS-B Signals using Distributed Public Sensors}
%
%\titlerunning{LoVe is in the Air}

\author{\IEEEauthorblockN{1\textsuperscript{st} Johanna Ansohn McDougall}
\IEEEauthorblockA{\textit{Security in Distributed Systems} \\
\textit{University of Hamburg}\\
Hamburg, Germany \\
johanna.ansohn.mcdougall@uni-hamburg.de}
\and
\IEEEauthorblockN{2\textsuperscript{nd} Alessandro Brighente}
\IEEEauthorblockA{\textit{Department of Mathematics} \\
\textit{University of Padova}\\
Padua, Italy \\
alessandro.brighente@unipd.it}
\and
\IEEEauthorblockN{3\textsuperscript{rd} Willi Großmann}
\IEEEauthorblockA{\textit{Helmut-Schmidt-University} \\
Hamburg, Germany \\
grossmann@hsu-hh.de}
\and
\IEEEauthorblockN{4\textsuperscript{th} Ben Ansohn McDougall}
\IEEEauthorblockA{\textit{Lübeck University of Applied Sciences} \\
Lübeck, Germany \\
ben.mcdougall@th-luebeck.de}
\and
\IEEEauthorblockN{5\textsuperscript{th} Joshua Stock}
\IEEEauthorblockA{\textit{Security in Distributed Systems}\\
\textit{University of Hamburg}\\
Hamburg, Germany \\
joshua.stock@uni-hamburg.de}
\and
\IEEEauthorblockN{6\textsuperscript{th} Hannes Federrath}
\IEEEauthorblockA{\textit{Security in Distributed Systems}\\
\textit{University of Hamburg}\\
Hamburg, Germany \\
hannes.federrath@uni-hamburg.de}
}
\maketitle              % typeset the header of the contribution
\begin{tikzpicture}[remember picture, overlay]
  \node[font=\sffamily\normalsize, yshift=-0.7cm, text centered, text width=\paperwidth, anchor=north west] at (current page.north west) {%
This paper has been accepted for publication at IEEE International Conference on Communications (ICC) 2023 %\url{TODO}
  };
\end{tikzpicture}
\begin{abstract}
The \ac{adsb} message scheme was designed without any authentication or encryption of messages in place. It is therefore easily possible to attack it, e.g., by injecting spoofed messages or modifying the transmitted \ac{gnss} coordinates. In order to verify the integrity of the received information, various methods have been suggested, such as multilateration, the use of Kalman filters, group certification, and many others. However, solutions based on modifications of the standard may be difficult and too slow to be implemented due to legal and regulatory issues. A vantage far less explored is the location verification using public sensor data. In this paper, we propose LoVe, a lightweight message verification approach that uses a geospatial indexing scheme to evaluate the trustworthiness of publicly deployed sensors and the \ac{adsb} messages they receive. With LoVe, new messages can be evaluated with respect to the plausibility of their reported coordinates in a location privacy-preserving manner, while using a data-driven and lightweight approach. By testing our approach on two open datasets, we show that LoVe achieves very low false positive rates (between 0 and 0.001\,06) and very low false negative rates (between 0.000\,65 and 0.003\,34) while providing a real-time compatible approach that scales well even with a large sensor set. Compared to currently existing approaches, LoVe neither requires a large number of sensors, nor for messages to be recorded by as many sensors as possible simultaneously in order to verify location claims. Furthermore, it can be directly applied to currently deployed systems  thus being backward compatible.        
\end{abstract}


\section{Introduction}\label{sec:introduction}
%\todo{Add comparison to other papers: Which kind of checks do they do?}
The \ac{adsb} protocol is a system designed to monitor aircraft positions via constantly transmitted messages. Indeed, aircraft use \ac{adsb} to communicate with ground control stations, other aircraft, or other \ac{adsb} enabled receivers~\cite{international2014ads}. While used as a secondary source of real-time air traffic monitoring to radar systems in metropolitan areas, it is often the single source of information in sparsely populated areas, difficult terrain and over oceans. Thanks to its broadcast nature, any \ac{adsb} enabled device can receive \ac{adsb} messages and obtain sensitive information on aircraft, such as their identifier and location. Also, \ac{cots} devices with limited resources can be used to receive and collect \ac{adsb} messages and, hence, aircraft information. This lead to the creation of the OpenSky Network~\cite{schafer2014bringing}, where aircraft are monitored in a crowd-sourced fashion. The possibility of collecting aircraft information is further facilitated by the fact that, by design, \ac{adsb} does not include any encryption or authentication measures~\cite{wu2020security}. While this provides fast access to aircraft data to be used for real-time monitoring, the lack of security measures provides attackers with the possibility for packet spoofing. In particular, an attacker might be able to create packets and inject false information on the presence of airplanes, or report false location claims. In fact, \ac{adsb} utilises the \ac{gnss} system to retrieve an aircraft's location, but does not encrypt or sign this information, which can hence be modified by an attacker. The consequences of these attacks range from mild (e.g., distraction on the flight deck or in the control room) to more severe such as violations of mandatory safety separations, and eventually aircraft collisions \cite{RIAHIMANESH201716}. The implementation of these attacks is not only an academic issue: it also captured the attention of a wider audience~\cite{greenberg2012next, ghostcostin2012}. Therefore, it is fundamental to develop solutions for real-time and scalable verification of \ac{adsb} messages.

%\begin{table*}
%\caption{Packet structure of ADS-B packets}
%\centering
%\begin{tabular}{|c|c|c|c|c|c|}
%\hline
%Preamble & Downlink Format & Capability & Aircraft Addr. & ADS-B Data & Parity Check\\ 
%\hline
%& 5 bit &  3 bit &  24 bit &  56 bit&  24 bit\\    
%\hline
%\end{tabular}
%\label{packet structure}
%\end{table*}

To verify the authenticity of \ac{adsb} messages, the literature proposes three different approaches. The first approach is to redesign the \ac{adsb} protocol itself to include cryptographic means to guarantee integrity and authentication. This could for example be done using public key-based, identity-based message encryption, or via Message Authentication Codes (MACs)~\cite{wu2020security}. However, \ac{adsb} is slow to be deployed due to different countries' regulations and the scale on which a modification has to be rolled out. Therefore, a change in protocol is unlikely to happen any time soon~\cite{MLforAttackDetectionJansen2021}.

The second approach %to verifying the authenticity of an \ac{adsb} signal 
entails fingerprinting the signals. This does not require a change in protocol; the idea is to perform software-based, hardware-based or channel-based fingerprinting~\cite{wu2020security}.

The third approach attempts to verify the signal origin: the use of \ac{mlat} has been suggested to approximate the legitimate origin, and additionally the use of Kalman filtering or group certification using signals received by other airplanes. While this system is very reliable, it is costly as it requires a gigantic amount of sensors to be deployed to receive accurate measurements: Using a test set of 8 sensors distributed within a 100\,km radius, Strohmeier et al.~\cite{strohmeier2015} show that MLAT is only applicable to 9.73\,\% of the messages due to the limited amount of sensors that record the same message. 
%In this field, several publication have recently attempted to provide verification of \ac{adsb} messages on distributed crowd sensors, e.g. using \ac{mlat} on the signals \cite{darabseh2020mavpro} (
Another \ac{mlat}-based approach to determine the trustworthiness of \ac{adsb} messages is MAVPro~\cite{darabseh2020mavpro}, a message verification protocol. The authors use \ac{mlat} in areas with large sensor coverage, and additionally rely on various other information like predicted trajectory information, preset anchors and information on flight tracks in areas with lower coverage.

Jansen et al.~\cite{MLforAttackDetectionJansen2021} propose a \ac{ml} approach utilizing vectorsets and evaluating \ac{adsb} sensor response patterns by applying a random-forest model. 
%Using a data set of \ac{adsb} messages collected via 729 sensors distributed over Europe in the OpenSky Network, they construct a vector containing the sensor state for every single one of the messages received on a specific day. This vector is representative of all sensors that recorded a particular message. They then apply the \ac{ml} technique decision trees (DT) to the vectors. This way, they are able to identify various attacks on the ADS-B system. 
%While the approach is applicable to a sensor set of size 729, we argue that it doesn't scale, e.g., in a world-wide scenario where thousands or hundred-thousands of sensors are in use. 
%However, even pattern recognition models cannot find characteristics in identical data samples. 
When evaluating \ac{adsb} data with machine learning, two problems arise. First, the feature space does not scale: Each sensor corresponds to a feature, which means that the feature space grows quadratically with increasing local coverage. In addition, an algorithm trained with data collected from a certain area cannot be transferred to other areas, because the feature space would change. Second, the chosen feature space is only slightly characteristic. Especially in areas with low sensor coverage, the necessary redundancy to generate characteristic patterns is missing.
%\subsection{Related Work}\label{sec:related work}
%An approach typical to location assessment that is also used in telecommunication networks is \ac{mlat}.  Here, a minimum of four receivers are required to pick up the same signal. %Signal features like the time of transmission or time of arrival are used to accurately track the location of the sender. 
 %Conventional \ac{mlat} is therefore rather applicable in limited airspaces, e.g. in the vicinity of airports, where it is also commonly used. 
%Several publications attempt to localise signal origins using \ac{mlat}-like approaches but typically apply modifications to reduce the required sensor set: Two approaches~\cite{strohmeier2015} 
%perform either location verification or location estimation using the measurement of the \ac{tdoa}. The authors collect a data set and analyse its verifiability with respect to \ac{mlat} and then compare it to their own \ac{tdoa}-based approach. They adapt their approach to work with a minimum of 2 sensors %receiving one message 
%and provide an improvement to conventional \ac{mlat} of 41\,\%. 
%Various other approaches typically concern \textit{data fusion} for verification of \ac{adsb} messages, combining the positioning via \ac{adsb} with localisation via primary or secondary radar or other current traffic control systems \cite{wu2020security, strohmeier2015}. %While the combination of various localisation techniques are typically expensive to deploy, they also defeat the original purpose of using \ac{adsb} for surveillance in remote areas, difficult terrain or over oceans. 

Other approaches of signal localisation are group localisation using a fixed set of aircraft as a group  \cite{Sampigethaya-group-localisation2011}, distance bounding \cite{RIAHIMANESH201716} or fingerprinting techniques on the physical layer \cite{danev-phys-layer-id-2012, wu2020security} or the use of Kalman filters to estimate the future trajectory of an aircraft via recent directional information~\cite{krozel-kalman-2004}.

In this paper, we propose Location Verification (LoVe), a simple yet effective approach based on data collected from public sensors for the verification of aircraft location claims sent via \ac{adsb} messages.
While our approach also uses distributed sensors, we 
choose to use an ML-free approach to verify signal authenticity: We take measurements on the range of coordinates a specific sensor can normally receive, and can thereby infer whether a newly received set of coordinates is within the range of typically received signals. We do this by mapping historic flight data into a representative map, through which current data can be evaluated with respect to plausibility of the signal. 
We are, to the best of our knowledge, the first to combine this with the use of a geospatial indexing system, resulting in a system that scales well even with a very high sensor count. Furthermore, it performs knowledge-based comparisons without resorting to costly ML-computations, while maintaining sensor location privacy. It achieves real-time compatible execution times while maintaining very low false positive rates between \num{0} and \num{0.00106} and very low false negative rates between
\num{0.00065} and \num{0.00334}. LoVe is also easily extendable and can be integrated into various existing schemes. 
To this end, we contribute the following:

\begin{itemize}
    \item We propose a lightweight, privacy-preserving classification of sensors to evaluate the plausibility of received messages.
    \item We implement the classification scheme by calculating the message distribution per sensor for each hexagon over the hexagonal geospacial indexing system H3.
    \item We test and evaluate the proposed method using real-world data provided by the OpenSky Network and Flightradar24 and multiple different hexagonal resolutions.
    \item We provide our implementation in a public repository\footnote{\url{https://github.com/heddha/LoVe}} for others to test and use. 
\end{itemize}
 
This paper is structured as follows: In \Cref{sec:background}, we first provide a background on ADS-B and the system and threat model. We then introduce our approach -- LoVe -- in \Cref{sec:LoVe}. This section contains our detailed approach, data acquisition, the experimental analysis and the verification evaluation. \Cref{sec:future work} then proposes future work and concludes the paper. %\todo{adapt to structure}

\section{Preliminaries and Models}\label{sec:background}

In this section, we provide background information on \ac{adsb} and describe the system and threat model.
\subsection{Background on ADS-B}
\ac{adsb} is a broadcasting system in which aircraft transmit status information every 0.5 seconds \cite{strohmeier2015}. \ac{adsb} packets are 112 bit to 272 bit long packets, depending on the link standard (UAT or Extended Squitter) \cite{RIAHIMANESH201716}. They consist of a preamble, the downlink format, capabilities, the aircraft address, an \ac{adsb} data field and a parity check. %,  the exact packet structure of which can be seen in \Cref{packet structure}. 
The \textit{\ac{adsb} Data} field contains information on altitude, speed, destination, the airplane ID and coordinates that are determined via \ac{gnss}.

A distinction is made between \textit{\ac{adsb} Out} devices, which are \ac{adsb} transmitters, and \textit{\ac{adsb} In}, the receiving side. Ground stations require \textit{\ac{adsb} In} functionality to receive the \ac{adsb} signals.  In European airspace, airplanes heavier than 5.7 tons or faster than 250 knots are required to transmit their \ac{adsb} data via \textit{\ac{adsb} Out}, while the use of \textit{\ac{adsb} In} for airplanes is voluntary \cite{adsb-mandatory2020}.

The \ac{adsb} broadcast system does not include a collision detection mechanism for messages, and neither any means of security to ensure confidentiality or integrity. It is therefore possible to spoof messages, inject content into transmissions and to jam communications \cite{wu2020security}.
%\ac{adsb} does, however, contain two privacy features. One is \textit{Limiting Aircraft Data Displayed} (LADD), by which the amount of aircraft data displayed in flight tracking services can be reduced. The other is \textit{Privacy ICAO Aircraft} (PIA), through which a temporary identifier can be requested to increase the privacy of a specific airplane \cite{adsb-privacyFAQ}.

%\section{Simulation Environment and Data Acquisition}

%\subsection{Simulation Environments}
%openScope and BlueSky are 2 open source ATC simulation environments which were also compared in chapter 3 of \url{https://www.diva-portal.org/smash/get/diva2:1452531/FULLTEXT01.pdf}. OpenScope was mainly written in JavaScript while BlueSky was solely written in Python. From the above source: "OpenScope has a built-in traffic generator which means the game can be played endlessly while BlueSky does not have a traffic generator. Instead, BlueSky has prepared scenarios that the user can play and the user may also add aircraft at any time of playing."

%In their paper \cite{ontologyBasedAnomalyDetectionNeal2022}, Neal et al. use the EuroScope tower simulator\footnote{\url{https://www.euroscope.hu/wp/tower-simulator/}}. It can only be installed on Windows and reaquites for either Lockheed Martin’s Prepar3D 1.4+ or Microsoft’s Flight Simulator X SP2 to be preinstalled. 

%There also seems to be an ADS-B simulator included in the windows app store\footnote{\url{https://apps.microsoft.com/store/detail/adsb-simulator/9NFFJL95W28M?hl=en-us&gl=US}}, but I wasn't able to install it, unfortunately. 

\subsection{System Model}\label{sysMod}

We consider a system where legitimate aircraft periodically send \ac{adsb} messages, which are then recorded by geographically distributed ground sensors. Aircraft use both their local information (e.g., identifier) and information from \ac{gnss} satellites to create packets containing air traffic monitoring information according to the \ac{adsb} standard. All ground sensors share their received messages in a crowdsourcing fashion, reporting their measurements to a central server. We assume that the central server implements our location verification approach to detect the presence of malicious aircraft with the aim of securing the air traffic control-based management operations.

\subsection{Attacker and Threat Model}\label{thMod}

We consider an attacker able to monitor and send \ac{adsb} messages via \ac{cots} equipment, e.g., a \ac{sdr}. We also assume that the attacker may control various stationary \ac{adsb} transmitters. Based on the Federal Aviation Administration measurements\footnote{\url{https://www.faa.gov/air\_traffic/technology/adsb}}, a strong enough \ac{adsb} signal can be captured by sensors up to a maximum radius of 240\,km. A stationary \ac{adsb} transmitter transmitting messages with forged coordinates to simulate an airplane would therefore be in a legitimate coordinate range of the airplane for about 41 minutes when simulating an airplane travelling at a speed of about 600-800\,km/h and at an altitude of 10.9\,km. During this time, the signals would constantly be received by the same sensors, while the transmitted coordinates would place the aircraft in an area of less and less confidence, i.e. the received coordinates are implausible with respect to the position of the sensor that recorded it.  %A message transmitted this way can be detected due to the constantly decreasing confidence level. 

Our threat model considers various attacks, which we %i.e., injection of \ac{adsb} messages, GPS-spoofing and attacks on sensors for \ac{adsb} messages. We 
elaborate on in the following.

\parag{Message Injection and Ghost Plane Injection}
As \ac{adsb} messages are neither authenticated nor encrypted, it is easily possible to send legitimate looking messages. The attacker can, on the one hand, inject messages that appear like those of nearby planes, but possibly with altered details like coordinates or altitude. This way, it may look like the existing airplane is diverging from its original path and heading in a different direction. On the other hand, the attacker can also inject messages that appear like signals sent by a legitimate looking but nonexistent aircraft, so called \textit{ghost plane}. Both modified messages and injected ghost planes would appear on surveillance monitors and could potentially disrupt normal flight procedures and can cause other airplanes to modify their course to avoid collisions unless a verification of origin is performed on the message. %Additionally, other airplanes capable of receiving \textit{\ac{adsb} In} messages can divert from their path to avoid collisions with non-existent. %Both can also interfere with and disrupt air traffic surveillance systems.

\begin{table*}[h!]
\caption{Size and amount of hexagons per resolution in the H3 geospacial index and respective amount of entries in the database}
\centering
%\scalebox{0.9}{
\begin{tabular}{|c|c|c|c|c|c|c|c|c|}
\hline
\multirow{2}{*}{Resolution} & No. of & Avg. Hexagon & \multicolumn{2}{c|}{Sensor-location-pairs} & \multicolumn{2}{c|}{Avg. No. of msg. per sensor-location-pair} & \multicolumn{2}{c|}{Test time (s) for \num{200000} entries} \\
\cline{4-9}
& Hexagons &  Area (km²) & OpenSky & FlightRadar & OpenSky & FlightRadar & OpenSky & FlightRadar\\
\hline
2 & \num{5870} & \num{86801.78} & \num{9963} & \num{26133} &  \num{208721.51} & \num{535.43} & \num{0.549} & \num{0.586}\\
3 & \num{41150} & \num{12393.43} & \num{32339} & \num{65220} & \num{64302.93} & \num{214.54} & \num{0.602} & \num{0.678}\\   
4 & \num{288110} & \num{1770.35} & \num{143143} & \num{223018} & \num{14527.38} & \num{62.74} & \num{0.717} & \num{0.806}\\
5 & \num{2016830} & \num{252.90} & \num{776226} & \num{758857} & \num{2678.98} & \num{18.44} & \num{1.350} & \num{1.322}\\
6 & \num{14117870} & \num{36.13} & \num{4442948} & \num{1948721} & \num{468.04} & \num{7.18} & \num{5.138} & \num{2.483} \\
7 & \num{98825150} & \num{5.16} & \num{25086048} & \num{3741523} & \num{82.89} & \num{3.74} & \num{29.049}  & \num{4.438}\\
%8 & 691,776,110 & 0.73 & & 6,153,072 & & 2.27\\
%9 & 4,842,432,830 & 0.11 & & 8,764,381 & & 1.60\\
\hline
\end{tabular}
%}
\label{tab:h3res}
\end{table*}

\parag{Location Spoofing}
Location spoofing of \ac{adsb} messages can be done in two different ways, i.e., i) via GNSS  or ii) via message modification. In the first case, programs such as \textit{gps-sdr-sim}\footnote{\url{https://github.com/osqzss/gps-sdr-sim}} can be used to spoof \ac{gnss} receivers using a low-cost \ac{sdr} like the HackRF and a current map of the satellite constellations. This way, a constellation containing up to 12 virtual satellites can be spoofed; these can then ``send'' spoofed signals which appear like legitimate signals and are used as the basis for the positional computations \cite{gps-spoofing-wikipedia}. 
%While the signal from the \ac{sdr} is unlikely to reach an airplane in-flight, this attack is plausible to be conducted by airplane passengers. 
The second means of location spoofing is to inject different coordinates into existing \ac{adsb} messages in a message modification attack.
\ac{adsb} messages contain a parity check of 24 bits, where an error correction of up to 5 bits is possible. To modify the message, the significant parts of the existing message are either \textit{overshadowed} by a stronger signal or modified via bit-flipping~\cite{strohmeier2013survey}. 
This way, an attacker could perform a virtual trajectory modification attack and thereby cause discrepancies between the actual positions and those sent via \ac{adsb} and therefore received by air traffic surveillance. 
% Needless to say, injecting modifications into existing messages requires not only for the bits of interest to be flipped, but also for manipulation of the parity check. 
%When a received message can not be corrected via error correction, it is dropped. This behavior can be exploited by an attacker to cause message deletion. While our system can not detect message deletion attacks, 
%Coordinates outside the legitimate coordinate range are rated with low plausibility would therefore be detected and flagged.

\parag{Attacks on Sensors}
The next threat considered in this paper is an attack using \ac{adsb} sensors. In this scenario, an attacker attempts to inject messages into the system either by adding an own sensor and inserting illegitimate messages through it, or by flooding particular sensors with a large number of bogus messages.

%-> write threat model (and what we have to defend against) and then build system model from that

\section{Location Verification - LoVe}\label{sec:LoVe}

While various approaches to location verification of \ac{adsb} signals have been proposed, (c.f. \Cref{sec:introduction})%\todo{check whether related work is still there once Alessandro's done merging texts}
, we suggest a mask-based approach that can verify whether specific coordinates received from a sensor are legitimately within the range of coordinates that the sensor is able to sense. %, and b) attest the plausibility and validity of the specific sensor by grading how many signals in the specific coordinate range have been received before. 

In order to evaluate the recorded coordinates with respect to signals received by the same and other sensors in the area, we chose to use the H3\footnote{\url{https://h3geo.org}} hexagonal geospacial indexing system.
It provides evenly distributed hexagons in various resolutions, spread over the whole world map. 
This approach is more applicable to the scenario than, e.g., a mercator projection or others, as the cell size is fixed and therefore delivers comparable results for each cell. Using H3, one can choose between 16 different resolutions\footnote{\url{https://h3geo.org/docs/core-library/restable/}} per cell. 
The resolution, or cell size, defines the total amount of hexagons and the individual cell size, ranging from approximately four million square kilometres to one square metre. 
With respect to \ac{adsb} signals, which often travel distances of a few hundred kilometres, several of the resolutions can be disregarded for being too big or too small. With our approach, we only focus on the resolutions 2-7 as shown in \Cref{tab:h3res}, as their dimensions are most fitting with respect of the propagation of ADS-B signals.

% Using a data set of all flights recorded on July 23rd, 2021, obtained from FlightRadar24 and OpenSky Network, we 

\begin{figure}
    \centering
    \includegraphics[scale=0.57]{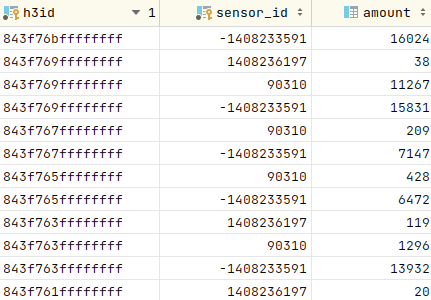}
    \caption{Sample data from the \textit{amount table} in resolution 4, sorted by h3id.}
    \label{fig:sample-amount-data}
\end{figure}

%\begin{table}
%    \centering
%\begin{tabular}{| L{0.3\columnwidth}  | R{0.3\columnwidth} | R{0.15\columnwidth}|}
%  \hline

%  H3ID & Sensor\_ID & Amount\\
%  \hline 
%  \hline
%  843f76bffffffff & -1408233591 & 16024\\
%  843f769ffffffff & 1408236197 & 38\\
%  843f769ffffffff & 90310 & 11267\\
%  843f769ffffffff & -1408233591 & 15831\\
%  843f767ffffffff & 90310 & 209\\
%  843f767ffffffff & -1408233591 & 7147\\
%  843f765ffffffff & 90310 & 428\\
%  843f765ffffffff & -1408233591 & 6472\\ 843f763ffffffff & 1408236197 & 119\\
%  843f763ffffffff & 90310 & 1296\\
%  843f763ffffffff & -1408233591 & 13932\\
%  843f761ffffffff & 1408236197 & 20\\
%  \hline
%\end{tabular}
%\end{table}

\begin{figure}
    \centering
    \includegraphics[scale=0.325]{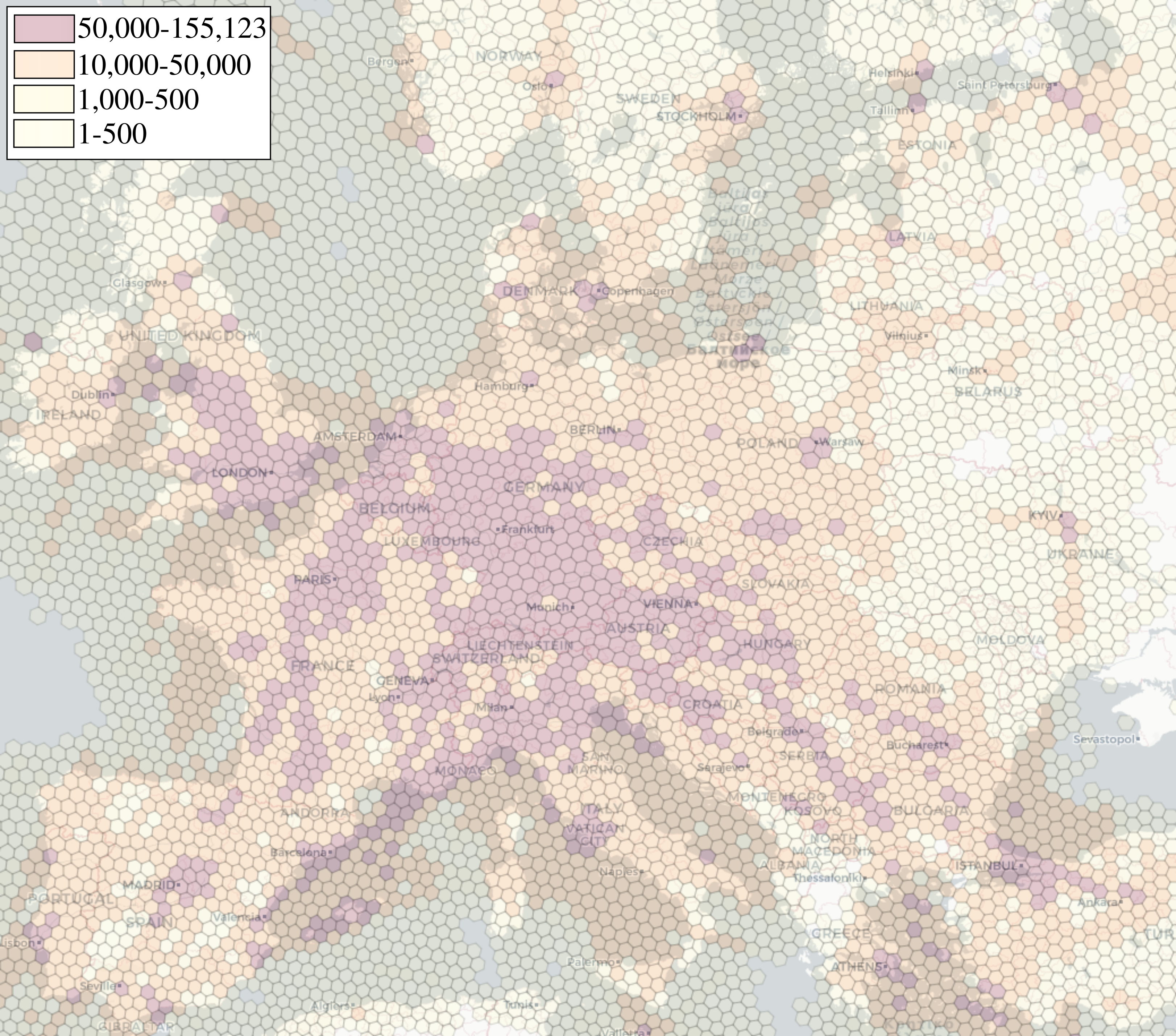}
    \caption{A map showing parts of Europe covered with H3 hexagons of resolution 4. The colours denote the amount of messages received by the sensor that received the most messages in the cell.}
    \label{fig:h34-map}
\end{figure}

The preparations of the LoVe-scheme were done as follows. 
We first acquire a dataset containing \ac{adsb} messages of a whole day (see \Cref{sec:data acquisition and structure} for details), process them, and feed them into a Postgres database, which then contains the sensors that recorded them, and the latitude and longitude that was contained within each message. 
We assume the data set recorded on that day is reliable and a trustworthy basis. From it, we construct one table, in the following called the \textit{amount table}, per resolution, as can be also observed in \cref{fig:sample-amount-data}: %, and under that assumptionable, can compute a list of all H3-representations with respect to the chosen resolution, in the following called H3ID, and the respective sensors in the cell, and additionally the amount of messages recorded per sensor. 
%Next, we construct two tables per resolution: 
It contains the \textit{h3id} (the H3-representation with respect to the chosen resolution), calculated from the latitude, longitude, and respective resolution, together with the sensor ID as the primary key and the respective amount of messages captured by the sensor within the area as the third column.
The total amount of sensor-location pairs per resolution for our data sets can be seen in \Cref{tab:h3res}, as well as the average number of messages per sensor-location pair.
Since the amount of cells rises the smaller the hexagon area is, higher resolutions cause a higher amount of sensor-location pairs as well. To illustrate the message distribution further, \Cref{fig:h34-map} shows a partial map of Europe distributed into H3-cells of resolution 4, with the colouring adapted to the maximum amount of messages captured by one sensor per hexagon.

%The second table, the \textit{homezone-table} contains one row per sensor and stores the \textit{home zone} of the sensor, i.e. the region where the sensor has received most messages. While one might argue that the actual sensor location could be in a completely different area and that the amount of signals are only representative of the amount of aircraft crossing the regions' airspace, we argue that this form of localisation of the sensor is sufficient while also location privacy-preserving, due to two factors: (1) The very high loss of \ac{adsb} messages with respect to the distance from the sensor: While \ac{adsb} messages already suffer a 30\,\% baseline message loss, every additional 10\,km introduce an additional percentage of message loss, reaching around 70\,\% at a distance of 300\,km  \cite{strohmeier2015}. (2) The amount of messages captured does not necessarily provide information about the exact sensor location, rather about the direction the antenna is directed towards. While it is likely that the actual sensor location is within a reasonable distance of the calculated \textit{home zone}, our calculation neither requires nor performs exact localisation. 

%The \textit{homezone-table} additionally holds the \textit{highest distance} value, which denotes how many cell hops are there between the home zone and the furthest recorded signal. 

\subsection{Verification Approach}
Our approach at origin verification of an incoming message $M$ with coordinates $lat$ and $long$ begins by calculating the respective $h3id$ for the two coordinates and the resolution. LoVe consists of the following two phases:
\begin{enumerate}
    \item Check whether the coordinates correspond to a $h3id$ in which the sensor has recorded messages before, using the \textit{amount-table}. If so, the signal is classified as legitimate.
    \item If not, check whether the specific sensor is known at all in our dataset.
\end{enumerate}

%we compile a list of the corresponding coordinates contained in the \ac{adsb} message and transform them to .

%The more signals were recorded in a specific cell, the higher we rate the reliability of the sensor in that area.

If a sensor then receives spoofed coordinates that are not within the range it typically receives coordinates in, we assume that the message is an illegitimate one.

\subsection{Data Acquisition and Structure}\label{sec:data acquisition and structure}

We test our approach using two different data sets recorded on the same day, July 23rd 2021: 
The first set was provided by the  OpenSkyNetwork\footnote{\url{https://openskynetwork.github.io/opensky-api/}}. They gave us access to their historical database, containing the stored \ac{adsb} messages from their crowdsourced network of sensors.
The second set was supplied by FlightRadar24\footnote{\url{https://www.flightradar24.com/}}, a commercial website offering live airtraffic monitoring.
To make our analysis comparable to others, we limit the range of both data sets to only contain entries from Europe, with the latitude within the boundaries of 30 and 75 and the longitude between -25 and 45.

To test our implementation, we required a labelled test set containing both false and true data. 
For the OpenSky-data, we chose the following approach to construct a test set: For legitimate flight data, we retrieved several hours worth of \ac{adsb} messages from the following day, July 24th, 2021 and limited the amount to \num{100000} randomly chosen true entries. We additionally generated false test data using the following approach: 
For every single sensor in the set, all coordinates in which a signal has ever been received are collected. From this set, we determine the minimum and maximum latitudes and longitudes it can legitimately receive information in. We then add or subtract a uniformly distributed random float between \num{0.1} and \num{10} to or from it, according to a random boolean\footnote{Notice that our approach in generating attacker data can not provide coherent spoofed values in successive time instants. This is not a problem, as our approach does not exploit any form of correlation or time dependency.}. 
While this is a very simple approach, it ensures that the coordinates are outside the range of expected coordinates and thereby simulates a \ac{gnss}-spoofing attack or an \ac{adsb} message injection attack using false coordinates. 
For the FlightRadar24-test set, we were unfortunately unable to acquire a second data set from the company. Instead, we randomly chose \num{100000} messages as true labelled test messages and removed them from the original set. For false labelled messages, we chose the same approach as described above for the OpenSky data set.

%In order to have a comparable set, we retrieved all flights recorded over Europe on July 23rd, 2021 as well, which amounted to 155,347,367 messages captured by 971 sensors. 

Both data sets are very different in content and structure, apart from the obvious differences in naming the sensors: The FlightRadar-set is first split up in flights surveilled and, for each flight, contains an additional CSV-file including all sensors that recorded the messages and for each message the received altitude, coordinates, velocity and others. 
Once % originally contained 58,491,973 \ac{adsb} messages recorded by 21,916, but after limiting it 
limited to the aforementioned coordinate boundaries, it contains \num{14092420} messages in total recorded by \num{11594} sensors. We additionally excluded another \num{100000} random messages from the set in order to construct a legitimate test set, which leaves us with \num{13992420} messages in total, recorded by \num{11582} sensors. 
On average, every sensor in this data set recorded approximately \num{1208.12} messages.  %From this record, we extracted all coordinates recorded per sensor and also fed them into a postgres database. % We then processed this information as explained in more detail in the following.
In this data set, only about 0.18\,\% of the messages were recorded by more than one sensor.

\begin{figure}
    \centering
    \includegraphics[clip, trim=0cm 0cm 0.1cm 0.3cm ,width=0.51\textwidth]{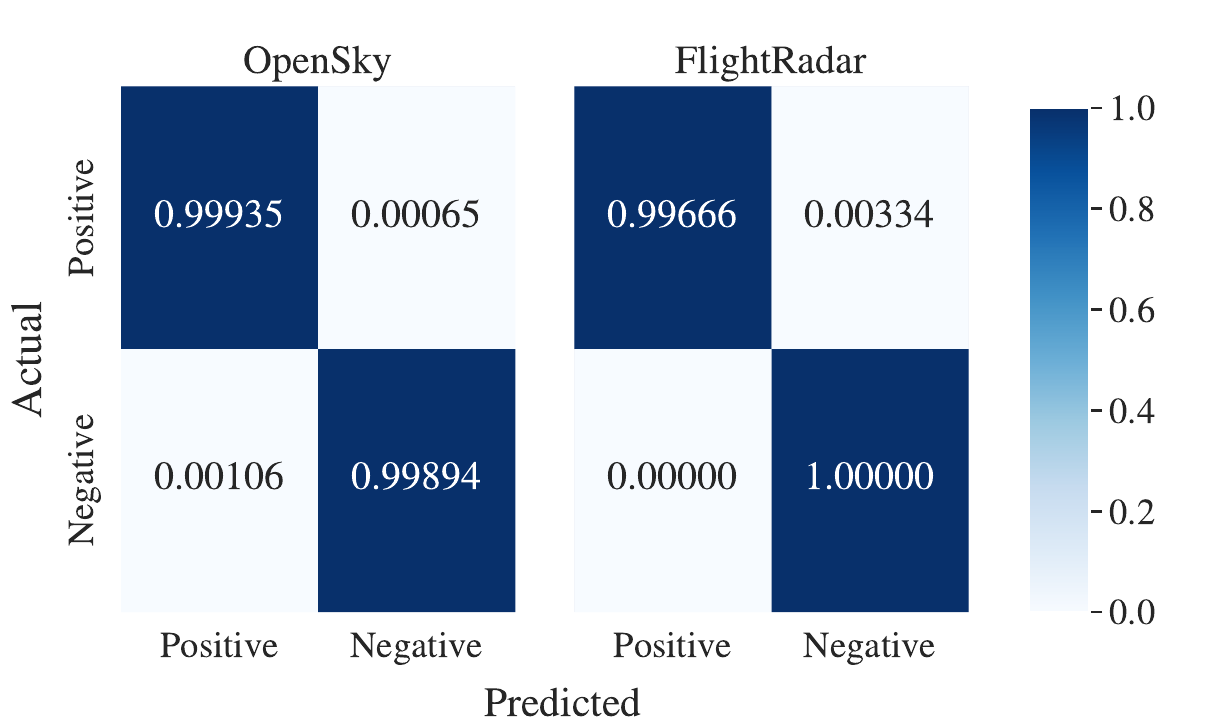}
    \caption{Comparison of the test runs using labelled test data sets of OpenSky (left) and FlightRadar (right), with our LoVe algorithm in resolution 4.}
    \label{fig:heatmap_res_4}
\end{figure}

The OpenSky data set contains far more messages recorded by far fewer sensors. %Per message recorded, it additionally contains information on overlapping sensors, e.g., instead of storing a single sensor per message, an array of multiple sensors that recorded the same message is stored with the message.  
This dataset contains \num{160526553} distinct \ac{adsb} messages recorded by \num{971} sensors. If the same message was recorded by different sensors, it was stored with an array containing all receiving sensors; the actual amount of single messages is therefore \num{2079492408}, while the list of sensors that captured one particular message has an average length of \num{12.95}. Every sensor recorded an average of \num{2141598.77} messages. %From this record, we extracted all coordinates recorded per sensor and fed both into a postgres database.

%Using the \ac{adsb}-messages 
%From these numbers, we can draw metrics specifying the meaningfulness and trustworthiness of sensors in specific hexagons: When choosing resolution 5, a sensor-location pair typically contains 27.95 messages. When we receive a new message, we can verify its meaningfulness by first converting the coordinates to the h3-hexagon ID. We then check for this hexagon-id which level of meaningfulness the sensor that sensed it has and test which other sensors should or could have received the same signal. 

For every resolution, our verification test is performed on a data set with which a reliable classification can be performed, as will be demonstrated in the following.

\subsection{Verification evaluation}\label{sec:verification evaluation}

To test our approach, we use both false and true labelled data, generated or acquired as described in \Cref{sec:data acquisition and structure}. Our evaluation achieves very low false positive rates between \num{0} and \num{0.00106} and very low false negative rates between \num{0.00065} and \num{0.00334} for both the OpenSky and FlightRadar datasets, as can be seen in \Cref{fig:heatmap_res_4}. The time required to test \num{200000} entries vary per resolution and span from \num{0.549} to \num{29.049} seconds for the OpenSky set and from \num{0.586} to \num{4.438} seconds for the FlightRadar set, as can also be seen in \Cref{tab:h3res}. The surprisingly high maximum execution time of the tests on the OpenSky dataset is directly linked to the amount of sensor-location-pairs of 25 million in resolution 7.

\begin{figure}
    \centering
    \includegraphics[clip, trim=0cm 0.15cm 0cm 0cm ,width=0.459\textwidth]{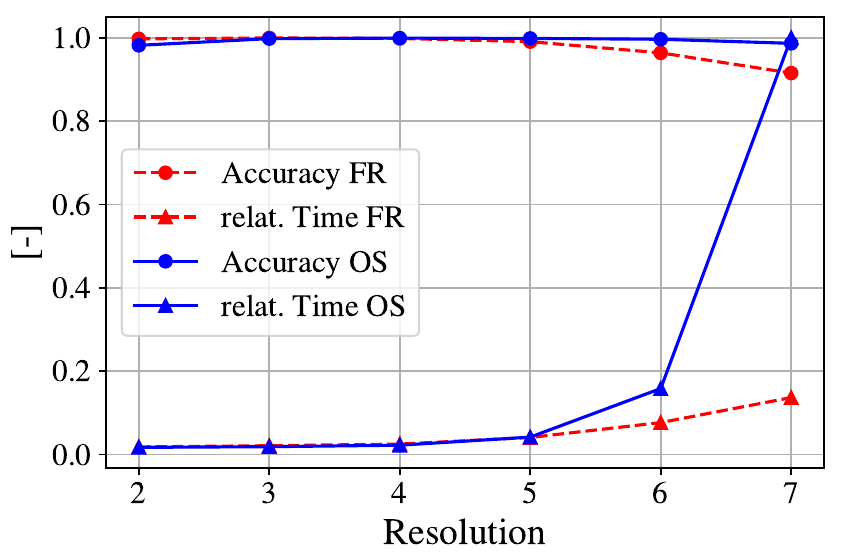}
    \caption{Accuracy and relative time needed to verify location accuracy using different resolutions. FR denotes results obtained with FlightRadar dataset, whereas OS those obtained with the OpenSky dataset.}
    \label{fig:acc_vs_res}
\end{figure}

The overall test results show that test results are best at resolution 4, while both higher and lower resolutions cause higher false negative and false positive rates. Larger table size simultaneously causes longer evaluation times. Both the relative execution time as well as the respective accuracy can be observed in \Cref{fig:acc_vs_res}. The comparison between the FlightRadar and OpenSky data sets also shows that our solution scales very well even with a much larger amount of sensors present, as can be observed in \Cref{tab:h3res}: even a data set containing 10 times the amount of sensors causes only slightly higher computation times, since our approach utilises the sensor-location-pairs as a basis for comparison and does not require to evaluate the message content.

Other approaches at location verification of \ac{adsb}-messages require multiple sensors to verify the message origin: Some attempt multilateration with a reduced amount of sensors \cite{darabseh2020mavpro, strohmeier2015}, others use \ac{ml} approaches that require a large sensor set \cite{MLforAttackDetectionJansen2021}. While their approaches only work with a large sensor set, our tests of the FlightRadar dataset show that our approach is also usable in the absence of a dense sensor network: The FlightRadar dataset does not contain any mentions on duplicate entries, e.g., single messages recorded by multiple sensors. To identify duplicates and see whether other data analysis strategies like those suggested in \cite{MLforAttackDetectionJansen2021} are applicable to this data set, we used the following approach: For every message, we calculated a hash over several entries of the ADS-B data field (altitude, heading, latitude, longitude, speed). %This revealed that %We then searched for duplicate hashes and identified that 
Only about \num{0.18}\,\% of the messages had duplicate hashes and had therefore  been recorded by more than one sensor. While the approaches mentioned above are inapplicable to such a data set, our approach is not impeded by it.

\subsection{Comparison to ML-Baseline}\label{sec:svm baseline}
To establish a speed comparison between the LoVe implementation and a machine learning approach, we chose to use a \ac{svm}, a well known, often used and easy to implement ML-classifier.
As in the LoVe implementation, we use batch learning mode, i.e. an entire set of of training examples used offline.
We use the same training data set as for the LoVe classifier but scale the features (sensor id, latitude and longitude) to a scale from 0 to 1.
We parameterize the \ac{svm} with values obtained from a hyperparameter search: C=10 and gamma=4641. Training the \ac{svm} took approximately 48h, since the training time of \ac{svm}s grows quadratically in the number of training samples. Afterwards, 
%After training the \ac{svm} classifier, 
 we  tested it against a FlightRadar data set with \num{200000} data records.
832 of these records were misclassified (all of those are false negatives); %which yields an accuracy of 99.584\% and a false negative rate of 0.00825135. 
compared to LoVe, the false negative rate is increased by 0.00491. The false positive rate and true negative rate stay at 0.0 respectively 1.0.

Testing the \num{200000} samples took another 91.3 minutes, which is demonstrably worse than LoVe.
While no training at all is needed for LoVe, the time to test the samples using SVM is increased by at least a factor of \num{6000}.

\section{Conclusion}\label{sec:future work}

The LoVe setup is very flexible, since we do not require to retrain models but only rely on minimal modifications, e.g. a sensor can easily be added or deleted from the database manually. We neither depend on formal groupings of the messages: While the OpenSky messages are grouped by the set of sensors  that received them, our approach works well with single messages that were observed only by a single sensor as well, as shown by the good results on the FlightRadar dataset.

% - Use of historical flight data for improved flight prediction
% - Check whether the airplane exists (icao)
% \todo[inline]{Alessandro, do you feel like writing something about this since it was your idea? If not, just delete it.}
Our suggested approach mainly consists of one database table per resolution, used to compare new \ac{adsb} messages to previously recorded ones, and verify their plausibility with respect to the sender that recorded it. The underlying mapping scheme that works as the basis for the sensor map makes use of the H3 hexagonal indexing system that provides intrinsic functions to modify the resolution and measure the distance between different cells. This way, we are able to compare various resolutions and find the best consensus between low false-negative rate and comparison speed. %Our approach solely requires information on the coordinates and sensor id with which a signal was received. 
%We explicitly avoid including various other information into the evaluation to establish a particularly lightweight and 
Additionally, our approach scales very well, both with respect to a high amount of sensors, messages and a larger surface area, due to the following constraints:
\begin{itemize}
    \item The number of possible cells is limited to a maximum for each resolution; 
    \item A large amount of sensors can be represented without significantly impacting the computational cost;
    \item With a large amount of messages, only the number of messages recorded per sensor and thereby the trustworthiness of the sensor is increased;
    \item Since we omit the message content, our system is particularly lightweight.% in comparison to costly and bloated ML-approaches.
\end{itemize}
LoVe is also privacy-preserving with respect to the location of the sensor: we don't require its position, but rather its reception area.
The amount of messages captured does not necessarily provide information about the  sensor location, rather about the direction the antenna is directed towards. While it is likely that the actual sensor location is within a reasonable distance of the area with most messages received, our calculation neither requires nor performs exact localisation.

Altogether, this makes our approach an easy-to-use basis that can be extended and integrated into already-established schemes.

%
% ---- Bibliography ----
%
% BibTeX users should specify bibliography style 'splncs04'.
% References will then be sorted and formatted in the correct style.
%
%\bibliographystyle{splncs04}
\bibliographystyle{IEEEtran}
%\raggedright
\typeout{}
\bibliography{sources}

\end{document}